\documentclass[galaxies,article,accept,oneauthors,pdftex,10pt,a4paper]{mdpi}
\firstpage{1}
\pubvolume{6}
\issuenum{2}
\articlenumber{65}
\pubyear{2018}
\copyrightyear{2018}
\history{Received: {23 April 2018}; Accepted: {15 June 2018}; Published: {19 June 2018}}
\updates{yes}
\usepackage{xspace}
\usepackage{graphicx,subfigure}
\usepackage{footnote}
\usepackage{tablefootnote}
\usepackage{sidecap}
\usepackage{booktabs}
\usepackage{multirow}
\usepackage{soul}
\usepackage{microtype}

\Title{Sliding along the Eddington Limit---Heavy-Weight Central Stars of Planetary Nebulae}

\Author{{Lisa L\" obling} $^{1,2}$\orcidA{} }

\address{%
$^{1}$ \quad Institute for Astronomy and Astrophysics, Kepler Center for Astro- and Particle Physics, \mbox{Eberhard Karls University}, Astronomy and Astrophysics, Sand 1, D-72076  T\" ubingen, Germany; loebling@astro.uni-tuebingen.de\\
$^{2}$ \quad European Southern Observatory (ESO), Karl-Schwarzschild-Stra\ss e 2, D-85748 Garching bei M\" unchen, Germany}

\abstract{Due to thermal pulses, asymptotic giant branch (AGB) stars experience periods of convective
mixing that provide ideal conditions for slow neutron-capture nucleosynthesis. These~processes are
affected by large uncertainties and are still not fully understood. By the lucky coincidence
that about a quarter of all post-AGB stars turn hydrogen-deficient in a final flash of the
helium-burning shell, they display nuclear processed material at the surface providing an
unique insight to nucleo\-synthesis and mixing.
We present results of non-local thermodynamic equilibrium spectral analyses of the extremely hot, hydrogen-deficient,
PG\,1159-type central stars of the Skull Nebula NGC\,246 and the ``Galactic Soccerballs'' Abell\,43 and NGC\,7094.}

\keyword{stars: abundances; stars: AGB and post-AGB; stars: atmospheres; stars: individual: WD~0044$-$121; stars: individual: WD 2134$+$25; stars: individual: WD 1751$+$106
}

\newcounter{Rco}
\newcommand{\ion}[2]{\textup{#1\,\textsc{\lowercase{#2}}}}

\newcommand{\Ionst}[1]{\setcounter{Rco}{#1}\Roman{Rco}}
\newcommand{\Ion}[2]{\mbox{#1\,{\scriptsize\Ionst{#2}}}}

\newcommand{\Jonw}[3]{\mbox{\ion{#1}{#2}~$\lambda\,#3$\,\AA}\xspace}
\newcommand{\Jonww}[3]{\mbox{\ion{#1}{#2}~$\lambda\lambda\,#3$\,\AA}\xspace}

\newcommand{\logg}{\mbox{$\log g$}\xspace}

\newcommand{\Teff}{\mbox{$T_\mathrm{eff}$}\xspace}

\newcommand{\pna}{A66\,43\xspace}
\newcommand{\wda}{WD\,1751$+$106\xspace}
\newcommand{\pnn}{NGC\,7094\xspace}
\newcommand{\wdn}{WD\,2134$-$125\xspace}
\newcommand{\pnnn}{NGC\,246\xspace}
\newcommand{\wdnn}{WD 0044$-$121\xspace}

\begin{document}

\def\aj{{AJ}}                   % Astronomical Journal
\def\actaa{{Acta Astron.}}      % Acta Astronomica
\def\araa{{ARA\&A}}             % Annual Review of Astron and Astrophys
\def\apj{{ApJ}}                 % Astrophysical Journal
\def\apjl{{ApJ}}                % Astrophysical Journal, Letters
\def\apjs{{ApJS}}               % Astrophysical Journal, Supplement
\def\ao{{Appl.~Opt.}}           % Applied Optics
\def\apss{{Ap\&SS}}             % Astrophysics and Space Science
\def\aap{{A\&A}}               % Astronomy and Astrophysics
\def\aapr{{A\&A~Rev.}}          % Astronomy and Astrophysics Reviews
\def\aaps{{A\&AS}}              % Astronomy and Astrophysics, Supplement
\def\azh{{AZh}}                 % Astronomicheskii Zhurnal
\def\baas{{BAAS}}               % Bulletin of the AAS
\def\bac{{Bull. astr. Inst. Czechosl.}}
% Bulletin of the Astronomical Institutes of Czechoslovakia
\def\caa{{Chinese Astron. Astrophys.}}
% Chinese Astronomy and Astrophysics
\def\cjaa{{Chinese J. Astron. Astrophys.}}
% Chinese Journal of Astronomy and Astrophysics
\def\icarus{{Icarus}}           % Icarus
\def\jcap{{J. Cosmology Astropart. Phys.}}
% Journal of Cosmology and Astroparticle Physics
\def\jrasc{{JRASC}}             % Journal of the RAS of Canada
\def\memras{{MmRAS}}            % Memoirs of the RAS
\def\mnras{{MNRAS}}             % Monthly Notices of the RAS
\def\na{{New A}}                % New Astronomy
\def\nar{{New A Rev.}}          % New Astronomy Review
\def\pra{{Phys.~Rev.~A}}        % Physical Review A: General Physics
\def\prb{{Phys.~Rev.~B}}        % Physical Review B: Solid State
\def\prc{{Phys.~Rev.~C}}        % Physical Review C
\def\prd{{Phys.~Rev.~D}}        % Physical Review D
\def\pre{{Phys.~Rev.~E}}        % Physical Review E
\def\prl{{Phys.~Rev.~Lett.}}    % Physical Review Letters
\def\pasa{{PASA}}               % Publications of the Astron. Soc. of Australia
\def\pasp{{PASP}}               % Publications of the ASP
\def\pasj{{PASJ}}               % Publications of the ASJ
\def\rmxaa{{Rev. Mexicana Astron. Astrofis.}}%
% Revista Mexicana de Astronomia y Astrofisica
\def\qjras{{QJRAS}}             % Quarterly Journal of the RAS
\def\skytel{{S\&T}}             % Sky and Telescope
\def\solphys{{Sol.~Phys.}}      % Solar Physics
\def\sovast{{Soviet~Ast.}}      % Soviet Astronomy
\def\ssr{{Space~Sci.~Rev.}}     % Space Science Reviews
\def\zap{{ZAp}}                 % Zeitschrift fuer Astrophysik
\def\nat{{Nature}}              % Nature
\def\iaucirc{{IAU~Circ.}}       % IAU Cirulars
\def\aplett{{Astrophys.~Lett.}} % Astrophysics Letters
\def\apspr{{Astrophys.~Space~Phys.~Res.}}
% Astrophysics Space Physics Research
\def\bain{{Bull.~Astron.~Inst.~Netherlands}}
% Bulletin Astronomical Institute of the Netherlands
\def\fcp{{Fund.~Cosmic~Phys.}}  % Fundamental Cosmic Physics
\def\gca{{Geochim.~Cosmochim.~Acta}}   % Geochimica Cosmochimica Acta
\def\grl{{Geophys.~Res.~Lett.}} % Geophysics Research Letters
\def\jcp{{J.~Chem.~Phys.}}      % Journal of Chemical Physics
\def\jgr{{J.~Geophys.~Res.}}    % Journal of Geophysics Research
\def\jqsrt{{J.~Quant.~Spec.~Radiat.~Transf.}}
% Journal of Quantitative Spectroscopy and Radiative Transfer
\def\memsai{{Mem.~Soc.~Astron.~Italiana}}
% Mem. Societa Astronomica Italiana
\def\nphysa{{Nucl.~Phys.~A}}   % Nuclear Physics A
\def\physrep{{Phys.~Rep.}}   % Physics Reports
\def\physscr{{Phys.~Scr}}   % Physica Scripta
\def\planss{{Planet.~Space~Sci.}}   % Planetary Space Science
\def\procspie{{Proc.~SPIE}}   % Proceedings of the SPIE

\let\astap=\aap
\let\apjlett=\apjl
\let\apjsupp=\apjs
\let\applopt=\ao

%%%%%%%%%%%%%%%%%%%%%%%%%%%%%%%%%%%%%%%%%%
%% Only for the journal Gels: Please place the Experimental Section after the Conclusions

%%%%%%%%%%%%%%%%%%%%%%%%%%%%%%%%%%%%%%%%%%
\setcounter{section}{0} %% Remove this when starting to work on the template.
%\vspace*{-1.5mm}
\section{Introduction}

\begin{figure}[H]
\centering
\hspace{-2.4em}\subfigure{\includegraphics[height=0.27\textheight]{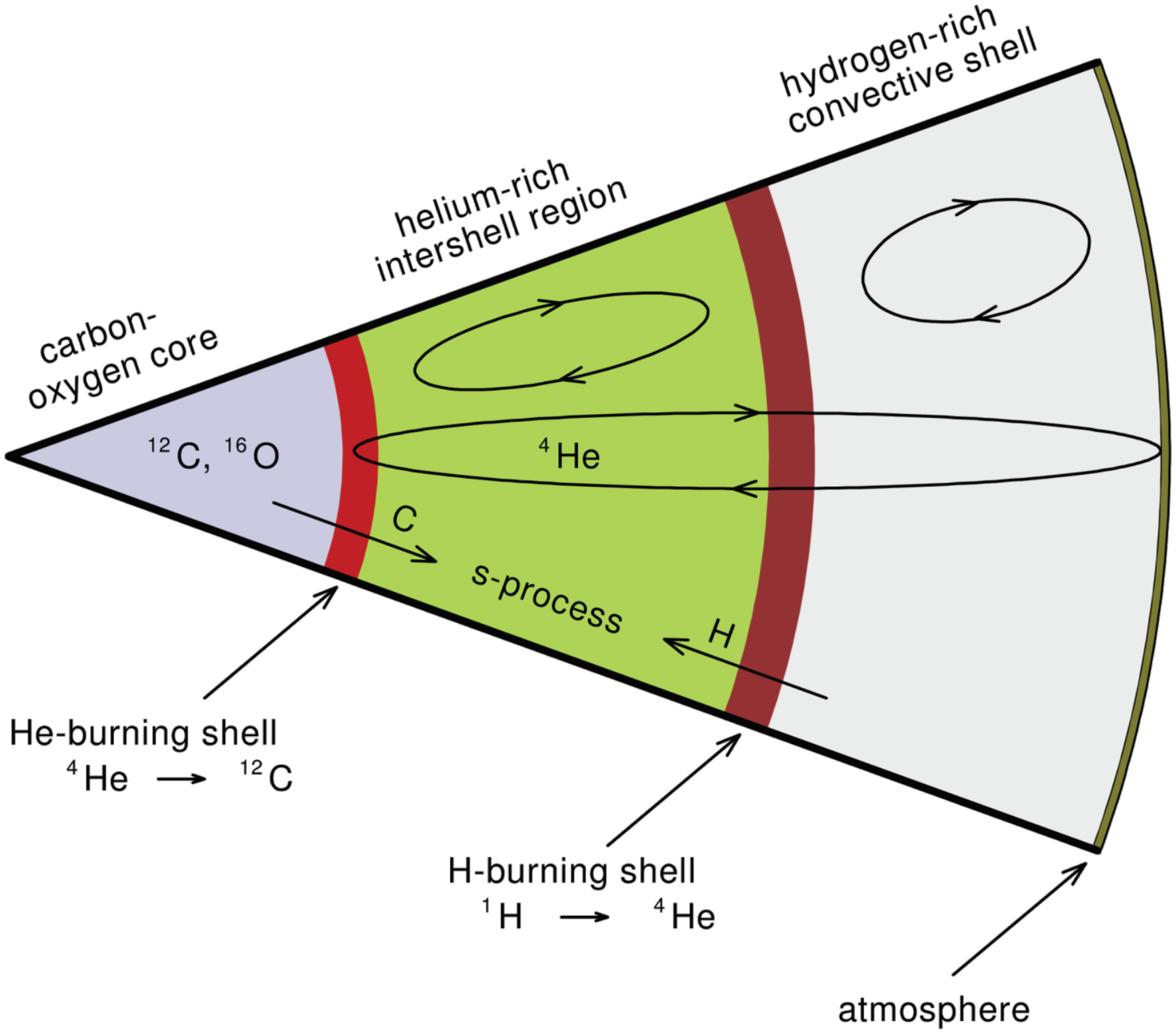}}
\subfigure{\includegraphics[height=0.27\textheight]{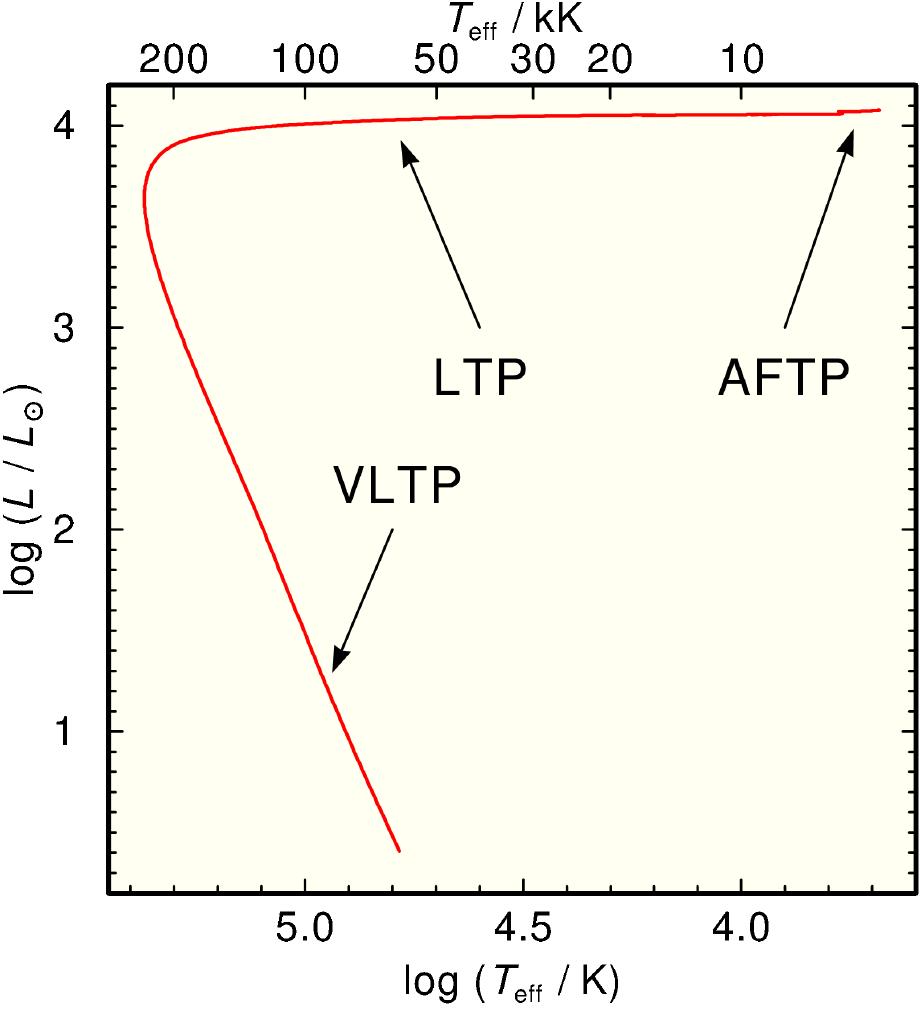}}
\vspace*{-4mm}
\caption{(\textbf{Left}): Internal structure of an AGB star (from \cite{rauchetal2008}); (\textbf{Right}): Occurrence of AFTP, LTP, and~VLTP in the Hertzsprung-Russell diagram.\vspace*{-4mm}}%please define AGB in figure.
\label{fig:intro}
\end{figure}

There are different evolutionary channels a star may go through after leaving the asymptotic giant branch (AGB). About a quarter become hydrogen (H) deficient as a result of a late flash of the helium (He)-burning shell (late thermal pulse, LTP, c.f., \cite{wernerherwig2006}).
For stars still located on the AGB at this event (AGB final thermal pulse, AFTP, Figure~\ref{fig:intro}), the H-rich envelope (with a mass of $10^{-2}\,M_\odot$) is mixed with the helium (He) rich intershell material ($10^{-2}\,M_\odot$). If the final flash happens after the departure from the AGB (envelope mass $\le 10^{-4}\,M_\odot$), H is either diluted by mixing in the LTP, if the nuclear fusion is still ``on'', or it is mixed down and totally consumed by the He-shell, if the star is already on the white dwarf cooling track, fusion is ``off'' and, thus, no entropy border exists any more. Predicted H mass fractions are 0.20 for an AFTP, 0.02 for an LTP and 0.00 for a VLTP (\cite{wernerherwig2006,herwig2005}, and references therein).

The three objects presented in this work belong to the spectral type of PG\,1159 stars (effective~temperatures of 75,000\,K\;$\leq$\;$T_\mathrm{eff}$\;$\leq$\;250,000\,K  and
surface gravities of 5.5\;$\leq$\;$\log (g\,/\,\mathrm{cm/s^2})$\;$\leq$\;8.0,~\cite{wernerherwig2006}) resulting from the H-deficient evolutionary channel.
The~central stars of the planetary nebulae (CSPNe) NGC\,7094 (PN G066.7$-$28.2 \cite{1992ackeretal}; CS: WD\,2134$-$125 \cite{1999mccooksion}) and Abell\,43 (PN G036.0$+$17.6 \cite{1992ackeretal}; WD\,1751$-$106~\cite{1999mccooksion}), known as spectroscopic twins, belong to the sub-type of hybrid PG\,1159-stars \citep{napiwotzki1999} exhibiting H lines in their spectra and, thus, resulting from an AFTP. In previous analyses, $\Teff = 100 \pm 15\,\mathrm{kK}$, $\log g = 5.5 \pm 0.2$, and a deficiency in Fe and Ni of $<$1\,dex were determined \cite{ziegleretal2009}. These stars are known to be fast rotators with rotational velocities of $v_\mathrm{rot} = 46 \pm 16\, \mathrm{km}/\mathrm{s}$ and $42 \pm 13\, \mathrm{km}/\mathrm{s}$, respectively \cite{rauchetal2004}.

\vspace{6pt}

\textls[-15]{WD\,0044$-$121 \cite{1999mccooksion} is the PG\,1159-type central star of the Skull Nebula (NGC\,246, PN~G118.8$-$74.7~\cite{1992ackeretal}) and with a mass of $0.7\,M_\odot$, it is among the most massive of this type. Previous~analyses revealed $\Teff = 150\,\mathrm{kK}$, $\log g = 5.7$, and a Fe and Ni deficiency of about 0.25\,dex. It rotates with $77 ^{+23}_{-17}\, \mathrm{km}/\mathrm{s}$ \cite{rauchetal2004}.
These stars are unique probes for AGB nucleosynthesis since they show the intershell material at the surface and exhibit a strong enough wind to prevent processes like gravitational settling and radiative levitaion to tamper the original composition. The aim of this analysis is to consider atomic data of trans-iron elements ($Z \geq 30$) that became available recently~(\cite{rauchetal2017}, and~references therein) for a NLTE stellar-atmosphere calculation to determine abundances for a large set of elements to draw conclusions on the evolutionary history of these stars and on AGB~nucleosynthesis.}
%The order of the section titles is: Introduction, Materials and Methods, Results, Discussion, Conclusions for these journals: aerospace,algorithms,antibodies,antioxidants,atmosphere,axioms,biomedicines,carbon,crystals,designs,diagnostics,environments,fermentation,fluids,forests,fractalfract,informatics,information,inventions,jfmk,jrfm,lubricants,neonatalscreening,neuroglia,particles,pharmaceutics,polymers,processes,technologies,viruses,vision
\vspace{2pt}
\section{Observations and Model Atmospheres}
%\vspace*{-1.5mm}

Our analysis is based on high-resolution observations from the far ultraviolet (FUV) to the optical wavelength range. The log for the observations retrieved from the Barbara A. Mikulski Archive for Space Telescopes (MAST) and the European Southern Observatory (ESO) Science Archive is shown in Table~\ref{tab:obslog}.
For the calculation of synthetic spectra, we employ the T\" ubingen NLTE Model Atmosphere Package (TMAP\footnote{\url{http://astro.uni-tuebingen.de/~TMAP}}, \cite{werneretal2003,werneretal2012,rauchdeetjen2003}) working under the assumption of a plane-parallel geometry and hydrostatic and radiative equilibrium. For WD\,2134$+$125 and WD\,1751$+$106, we consider opacities of 31 elements from H to barium Ba. The models for WD\,0044$-$121 include the elements H, He, C, N, O, F, Ne, Mg, Ar, Ca, Fe, and Ni. To analyse transiron-element abundances, we added them individually in line-formation calculations in which the temperature structure is kept fixed and the occupation numbers for the levels are calculated. We performed test calculations to confirm that these elements do not affect the atmospheric structure (see also (\cite{rauchetal2017}, and references therein)).
Atomic data for H, He, and the light metals (atomic weight $Z < 20$) is retrieved from the T\" ubingen Model Atom Database (TMAD\footnote{\url{http://astro.uni-tuebingen.de/~TMAD}}, \cite{rauchdeetjen2003}), for the iron group (Ca-Ni) elements, we used Kurucz's line lists\footnote{\url{http://kurucz.harvard.edu/atoms.html}} \cite{kurucz2009,kurucz2011}, and~the transiron-element data is available via the T\" ubingen Oscillator Strength Service (TOSS\footnote{\url{http://dc.g-vo.org/TOSS}}). For~all elements with $Z \geq 20$, a statistical approach is used based on super lines and super levels \cite{rauchdeetjen2003}.
\begin{table}[H]
\centering
\caption{Observation log for WD\,2134$+$125, WD\,1751$+$106, and WD\,0044$-$121.}
\label{tab:obslog}
\begin{tabular}{cccccc}
\toprule
\textbf{Object} & \textbf{Instrument }& \textbf{Dataset/ Prog. ID} & \textbf{Start Time (UT)} & \textbf{Exp. Time (s)}\\
\midrule
WD\,2134$+$125 & FUSE\,$^\mathrm{a}$ & P1043701000 & 2000-11-13 08:53:28  & 22,754\\
& STIS\,$^\mathrm{b}$  & O8MU02010   & 2004-06-24 20:43:30  & $650$\\
& STIS & O8MU02020   & 2004-06-24 22:19:29 & $656$\\
& STIS & O8MU02030   & 2004-06-24 23:55:29 & $655$\\
& UVES\,$^\mathrm{d}$ & 167.D$-$0407(A) & 2001-08-21 02:00:03  & $300$\\
& UVES & 167.D$-$0407(A) & 2001-09-20 01:40:00 & $300$\\
WD\,1751$+$106 & FUSE & B0520201000 & 2001-07-29 20:41:47  & 11,438\\
& FUSE & B0520202000 & 2001-08-03 22:18:20  & $9528$\\
& GHRS\,$^\mathrm{c}$  & Z3GW0304T   & 1996-09-08 07:00:34  & $4243$\\
& UVES & 167.D$-$0407(A) & 2001-06-18 05:03:38  & $300$\\
& UVES & 167.D$-$0407(A) & 2001-07-26 01:27:48  & $300$\\
WD\,0044$-$121 & FUSE & E1180201000 & 2004-07-12 17:01:47  & $6505$\\
& STIS & O8O701010   & 2004-05-28 10:11:51  & $1967$\\
& STIS & O8O701020   & 2004-05-28 11:31:56 & $2736$\\
& UVES & 167.D$-$0407(A) & 2002-09-06 09:32:22  & $300$\\
& UVES & 165.H$-$0588(A) & 2000-12-07 02:26:29  & $300$\\
\bottomrule
\end{tabular}\\
\begin{tabular}{@{}c@{}}
\multicolumn{1}{p{\textwidth -.88in}}{\footnotesize a: Far Ultraviolet Spectroscopic Explorer; b: Space Telescope Imaging Spectrograph; c:  Goddard High Resolution Spectrograph; d: UV-Visual Echelle Spectrograph.}
\end{tabular}
\end{table}

%%%%%%%%%%%%%%%%%%%%%%%%%%%%%%%%%%%%%%%%%%
\section{Preliminary Results}
%\vspace*{-1.5mm}
For accurate abundance determinations, the precise knowledge of the stellar parameters is an inevitable requirement. We redetermined the surface gravity using our best fit for the observed line wings and depth increments of \Jonww{He}{ii}{4100.1, 4338.7, 4859.3, 5411.5} and \Jonww{H}{i}{4101.7, 4340.5, 4861.3}.
The temperature determination is based on the ionization equilibrium of \Ion{O}{5}/\Ion{O}{6} using \Jonw{O}{v}{1371.3} and \mbox{\textup{O\,\textsc{\lowercase{vi}}}~$\lambda\lambda\,1080.6, 1081.2,1122.3, 1122.6, 1124.7, 1124.9, 1126.3$}\xspace, \mbox{$1290.1, 1290.2, 1291.8, 1291.9$}\xspace. The redetermination of $v_\mathrm{rot}$ results in lower values for both hybrid PG\,1159 stars. The new values are summarized in Table~\ref{tab:finab}.
\begin{table}[H]
\caption{Parameters of \wdn, \wda, and \wdnn.}
\label{tab:finab}
\centering
\begin{tabular}{rccc}
\toprule
\noalign{\smallskip}
&\textbf{\wdn} & \textbf{\wda} & \textbf{\wdnn} \\
\midrule\noalign{\smallskip}
$T_\mathrm{eff}$\,/\,kK      & $115\pm 5$ & $115\pm 5$ &  $150\pm 10$              \\
$\log\,(g$\,/\,cm/s$^2$)    & $5.4\pm 0.1$ & $5.5\pm 0.1$  & $5.7\pm 0.1$     \\
$v_\mathrm{rot}$\,/\,km/s    & $28\pm 5$& $18\pm 5$& $75\pm 15$\\
$d$\,/\,kpc                 & $2.38^{+0.59}_{-0.70}$ & $2.94^{+0.82}_{-0.91}$ & $1.08^{+0.22}_{-0.26}$ \\
$M\,/\,M_\odot$             &  $0.64^{+0.06}_{-0.07}$ &  $0.60^{+0.09}_{-0.06}$&  $0.74^{+0.19}_{-0.23}$\\
$M_\mathrm{ini}\,/\,M_\odot$             &  $3.30^{+0.65}_{-1.43}$ &  $2.87^{+0.71}_{-2.14}$&  $3.91^{+2.55}_{-0.88}$\\
$\log\ ( L\,/\,L_\odot )$   &  $4.04^{+0.32}_{-0.33}$ &  $3.91^{+0.34}_{-0.32}$ &  $4.27^{+0.41}_{-0.59}$\\
%$R_\mathrm{PN}$\,/\,pc      &  $0.61^{+0.15}_{-0.18}$&$0.63^{+0.17}_{-0.19}$&$0.59^{+0.12}_{-0.14}$\\
%Exp. time\,/\,$10^3$\,a     &  $15.7^{+3.9}_{-4.6}$&$15.3^{+4.3}_{-4.7}$&$14.9^{+3.0}_{-3.6}$\\
\noalign{\smallskip}\bottomrule
\end{tabular}
\end{table}

\begin{figure}[H]
\centering
\hspace{-0.2em}\includegraphics[width=0.95\textwidth]{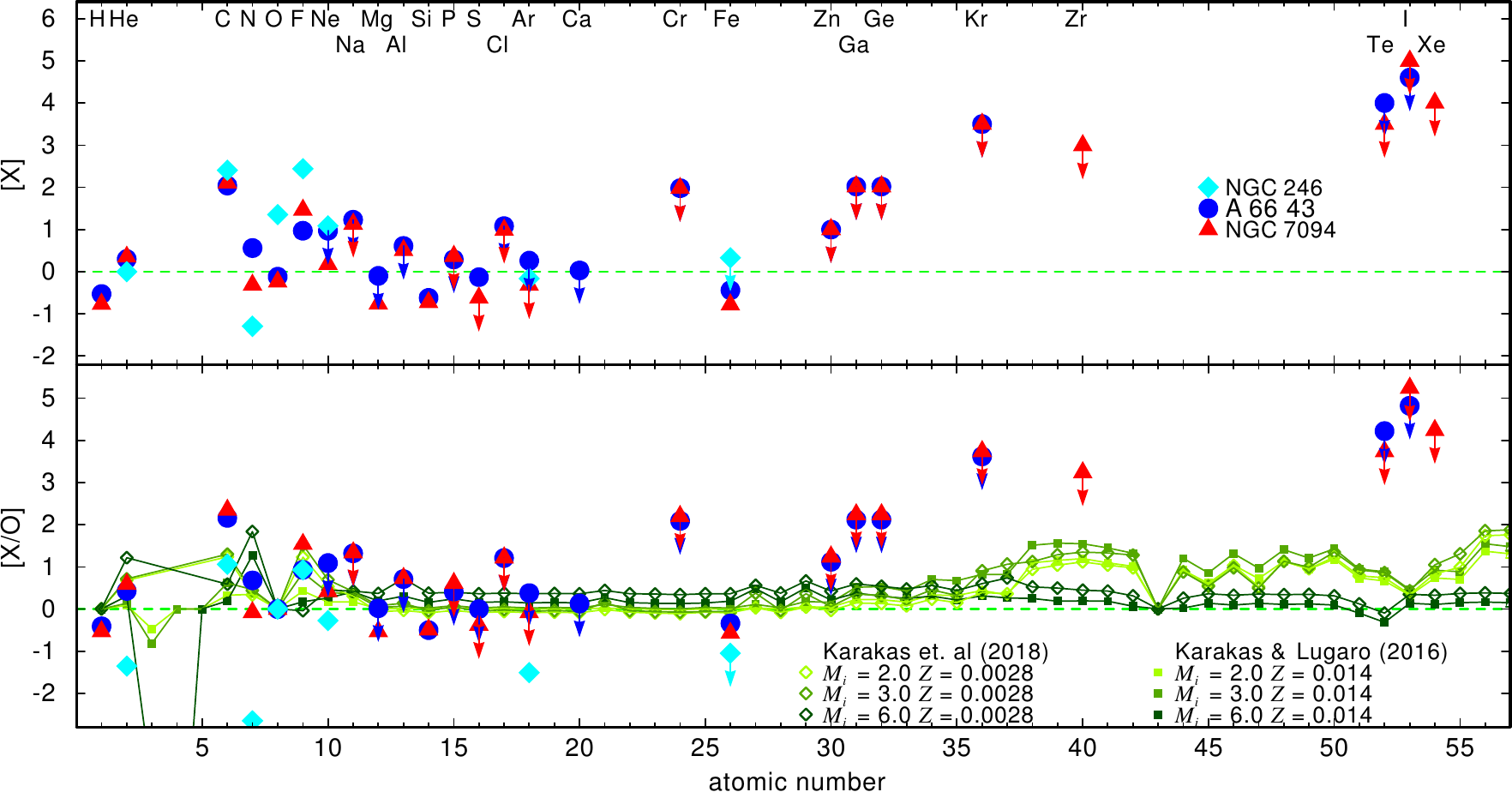}
\caption{Photospheric abundances (estimated errors $+$/$-$ 0.1 dex) of \wdn, \wda, and \wdnn. Arrows indicate upper limits. In the lower panel, the abundances in $[\mathrm{X}/\mathrm{O}]~=~\log \left(\mathrm{X} / \mathrm{O}\right)_\mathrm{surf} - \log \left(\mathrm{X} / \mathrm{O}\right)_\mathrm{solar}$ with the number fractions of element X and O are compared to the yields of the models from \cite{karakas2016,karakasetal2018} for different initial masses.\vspace*{-0.1cm}}
\label{fig:abund}
\end{figure}

\begin{figure}[H]
\centering
\includegraphics[width=0.60\textwidth]{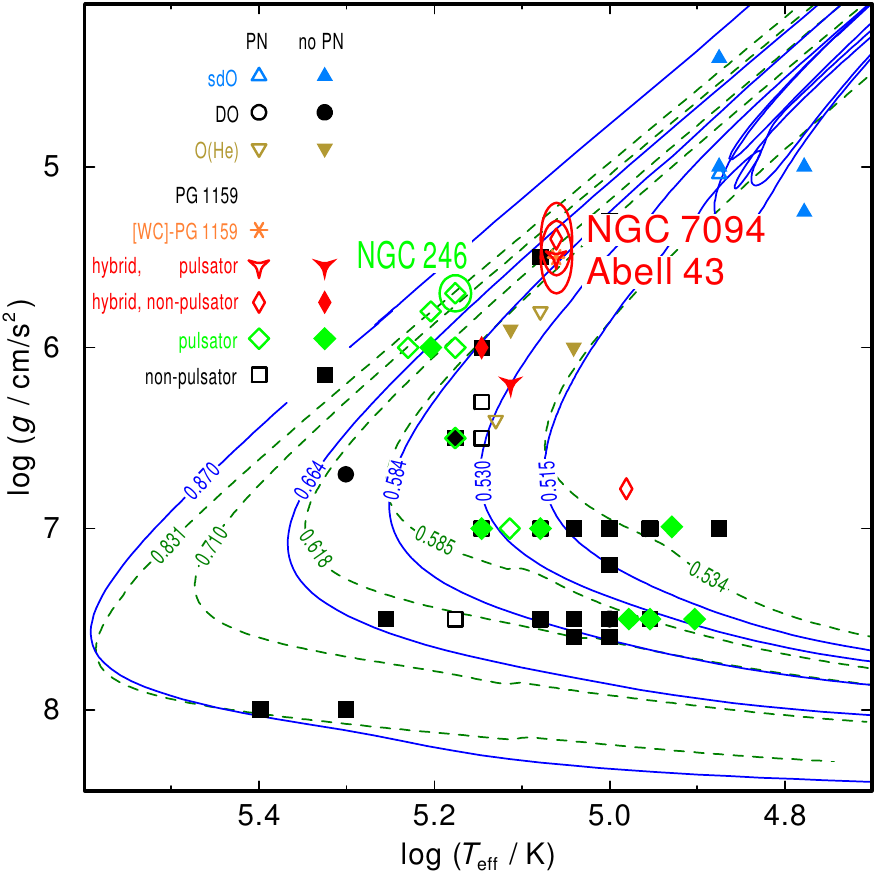}
%\vspace*{-1.5cm}
\caption{Positions of the CSPNe of \pnn , \pna , and \pnnn (within their error ellipses) and related objects in the $\log \Teff - \logg$ plane compared with evolutionary tracks (labeled with the respective masses in $M_\odot$) of VLTP stars (\cite{millerbertolamialthaus2006}, full lines) and of hydrogen-burning post-AGB stars (\cite{millerbertolami2016}, calculated with initial solar metalicity; dashed lines). \vspace*{-0.1cm}}\label{fig:tg}
\end{figure}

We found that the Fe abundance for both hybrid PG\,1159 stars needed to be corrected. The~lines of \Ion{Fe}{7} and \Ion{Fe}{8} appear weaker in the synthetic spectra due to the higher \Teff and rotational broadening that was not taken into account in previous works with the result that we find Fe to be less deficient ($[\mathrm{Fe}] =-0.79$ for \wdn and $[\mathrm{Fe}]= -0.43$ for \wda, $[\mathrm{X}] = \log$~(mass fraction/solar~mass fraction)). Due to the challenge of its fast rotation, no unambiguous identification of Fe lines was possible for \wdnn. Based on \Jonw{Fe}{vii}{1141.4} and \Jonww{Fe}{viii}{1006.1, 1148.2} an upper limit of the super-solar value $[\mathrm{Fe}]= 0.33$ is reasonable. No clear identification of trans-iron element lines was possible for any of the three stars. For the two hybrid PG\,1159 stars, the strongest computed lines were used to determine upper abundance limits for Zn, Ga, Ge, Kr, Zr, Te, I, and Xe. This was not possible for \wdnn, since the atomic data for the most prominent ionization stages of these elements in this high temperature range is still lacking. The abundance determinations are summarized in Figure~\ref{fig:abund}. Mass and luminosity were determined using He-burning post-AGB tracks~\cite{millerbertolamialthaus2006} (Figure~\ref{fig:tg}).
Using the flux calibration of  \cite{heberetal1984}, we find spectroscopic distances for the stars. The~values are summarized in Table~\ref{tab:finab}.
\vspace{6pt}

\unskip

%%%%%%%%%%%%%%%%%%%%%%%%%%%%%%%%%%%%%%%%%%
\section{Discussion}
%\vspace*{-1.5mm}

In our detailed spectroscopic analysis, we could confirm the high \Teff and low \logg of \wdnn which places this star close to the Eddington limit for PG\,1159 stars (Figure~\ref{fig:tg}) and approves its classification as one of the heaviest stars of this type. In return, this causes difficulties due to instabilities in calculating model atmospheres for this star. As improvement, models including wind effects will be employed in a further analysis of all these stars.
The Fe deficiency is not explained by nucleosynthesis models \cite{karakas2016,karakasetal2018} that predict solar Fe abundances. A speculative reason for low Fe abundances is the conversion into Ni and heavier elements due to neutron capture \cite{herwigetal2003}. An enhanced $\mathrm{Ni} / \mathrm{Fe}$ ratio or a clear enhancement of trans-iron elements would indicate this. Unfortunately, no~lines of these elements were identified. Given the high distances above and below the galactic plane, the progenitors of these stars could also belong to a low metallicity halo population. We included the low metallicity models for the Small Magellanic Cloud \cite{karakasetal2018} in Figure~\ref{tab:finab} for comparison but they do not reproduce the negative $[\mathrm{Fe}/\mathrm{O}]$ values for these stars and therefore cannot be consulted to explain the Fe deficiency.
Parallaxes of $(0.431 \pm 0.061)\,''$ and $(0.615 \pm 0.059)\,''$ for \wda and \wdn, respectively, were published in the second data release of the Gaia mission from which we can derive distances of $(2.32 \pm \, 0.33)\,$\,kpc and $(1.63 \pm 0.16)$\,kpc. The values lie below our spectroscopically measured distances but for \wda, the values still agree within the error limits. This might lead to the speculation of potentially too high surface gravity values.
%%%%%%%%%%%%%%%%%%%%%%%%%%%%%%%%%%%%%%%%%%
%%%%%%%%%%%%%%%%%%%%%%%%%%%%%%%%%%%%%%%%%%

\section{Conclusions}
%\vspace*{-1.5mm}

In our NLTE spectral analysis of the CSPNe \wdn, \wda, and \wdnn, we improved the determination of the stellar parameters and found, for the first two objects, upper~abundance limits for s-process elements that have never been analysed in any of these stars before. \wdnn shows no clue for s-process element enhancement which leads to the speculation that the s-process becomes less effective for higher mass and metallicity. \mbox{Unfortunately, the abundance} limits cannot be used to constrain the nucleosynthesis models. In a forthcoming analysis, we~plan to search for the radioactive element technetium and for s-process signatures in the ejected nebula~material.

%%%%%%%%%%%%%%%%%%%%%%%%%%%%%%%%%%%%%%%%%%
%%%%%%%%%%%%%%%%%%%%%%%%%%%%%%%%%%%%%%%%%%
\vspace{6pt}

%%%%%%%%%%%%%%%%%%%%%%%%%%%%%%%%%%%%%%%%%%
\funding{L.L. is supported by the German Research Foundation (DFG, grant WE\,1312/49-1) and by the ESO studentship programme.
The GAVO project had been supported by the Federal Ministry of Education and Research (BMBF)
at T\"ubingen (05\,AC\,6\,VTB, 05\,AC\,11\,VTB).
}
%%%%%%%%%%%%%%%%%%%%%%%%%%%%%%%%%%%%%%%%%%
\acknowledgments{The author thanks the reviewers for there fruitful comments and remarks to improve this proceedings paper. Sincere thanks are given to the SOC for awarding the follow-up price for the Lyra award for the best oral contribution of an early career scientist to this conference contribution.
This research has made use of
NASA's Astrophysics Data System and
the SIMBAD database, operated at CDS, Strasbourg, France.
}
\conflictsofinterest{The author declares no conflicts of interest.}

%%%%%%%%%%%%%%%%%%%%%%%%%%%%%%%%%%%%%%%%%%
%\vspace*{-0.5mm}

%=====================================
% References, variant A: internal bibliography
%=====================================
\reftitle{{References}}%ref 21 has not reffered to within the maintext. please confirm and revise.
%\vspace*{-1.5mm}

%%%%%%%%%%%%%%%%%%%%%%%%%%%%%%%%%%%%%%%%%%

%%%%%%%%%%%%%%%%%%%%%%%%%%%%%%%%%%%%%%%%%%
\end{document}